\documentclass[aps,prl,twocolumn,superscriptaddress,showpacs,showkeys,amssymb,amsmath]{revtex4-1}
\usepackage{graphicx}

\usepackage[T1]{fontenc}

\usepackage{color}

\newcommand{\V}[1]{\mathbf{#1}} 
 
\newcommand\Alfven{Alfv\'en }

\begin{document}

\title{A Majority of Solar Wind Intervals Support Ion-Driven Instabilities}

\author{K.~G. Klein}
\email[]{kgklein@email.arizona.edu}
\affiliation{Climate and Space Sciences and Engineering, University of Michigan, Ann Arbor, MI 48109, USA}
\affiliation{Lunar and Planetary Laboratory, University of Arizona, Tucson, AZ 85719, USA}
\author{B.~L. Alterman}
\affiliation{Climate and Space Sciences and Engineering, University of Michigan, Ann Arbor, MI 48109, USA}
\author{M.~L. Stevens}
\affiliation{Smithsonian Astrophysical Observatory, Cambridge, MA 02138 USA}
\author{D. Vech}
\affiliation{Climate and Space Sciences and Engineering, University of Michigan, Ann Arbor, MI 48109, USA}
\author{J.~C. Kasper}
\affiliation{Climate and Space Sciences and Engineering, University of Michigan, Ann Arbor, MI 48109, USA}
\affiliation{Smithsonian Astrophysical Observatory, Cambridge, MA 02138 USA}

\date{\today}

\begin{abstract}
We perform a statistical assessment of solar wind stability at 1 AU
against ion sources of free energy using Nyquist's instability
criterion. In contrast to typically employed threshold models which
consider a single free-energy source, this method includes the effects
of proton and He$^{2+}$ temperature anisotropy with respect to the
background magnetic field as well as relative drifts between the
proton core, proton beam, and He$^{2+}$ components on stability. Of
309 randomly selected spectra from the Wind spacecraft, $53.7\%$ are
unstable when the ion components are modeled as drifting
bi-Maxwellians; only $4.5\%$ of the spectra are unstable to
long-wavelength instabilities. A majority of the instabilities occur
for spectra where a proton beam is resolved. Nearly all observed
instabilities have growth rates $\gamma$ slower than instrumental and
ion-kinetic-scale timescales. Unstable spectra are associated with
relatively-large He$^{2+}$ drift speeds {and/or} a departure of
the core proton temperature from isotropy; other parametric
dependencies of unstable spectra are also identified.
\end{abstract}


\maketitle 

\emph{Introduction.}--- Plasma instabilities, wave-particle
interactions driven by departures from local thermodynamic
equilibrium, influence the dynamics of nearly collisionless systems,
including those frequently encountered in space and astrophysical
contexts.  In order to transfer free energy from plasma particles to
electromagnetic fields and drive unstable growth, non-equilibrium
attributes--- including anisotropic temperatures relative to the local
mean magnetic field, relative drifts between component distributions,
and more general agyrotropic features--- must either contribute to
sufficiently large departures from equilibrium or enable a resonant
interaction between fields and velocity-space structure in the
particle distribution. The determination of these conditions is
complicated in systems with many sources of free energy.

The large number of in situ observations of the solar wind, a
nearly-collisionless, low-density, high-temperature plasma emanating
from the Sun's surface, enables the statistical study of plasma
processes, including instabilities. Typical instability studies focus
on what unstable modes may arise due to a single free-energy source in
a reduced parameter space. As an example, the departure of the proton
temperature ratio $T_{\perp p}/T_{\parallel p}$ from isotropy, where
$\perp$ and $\parallel$ are defined with respect to the mean magnetic
field $\V{B}$, can drive \Alfven ion
cyclotron\cite{Kennel:1966,Davidson:1975},
mirror\cite{Tajiri:1967,Southwood:1993,Kivelson:1996}, parallel
firehose\cite{Quest:1996,Gary:1998}, \Alfven (or oblique)
firehose\cite{Hellinger:2000}, or CGL (or long-wavelength)
firehose\cite{Chew:1956} instabilities. Similar instabilities arise
for electron and minor ion temperature anisotropies, and other
instabilities arise due to drifts between the distributions. A recent
review of kinetic plasma instabilities can be found in Yoon
2017\cite{Yoon:2017}.

For each kind of unstable mode, one can determine using linear theory
the threshold value of a single parameter, assuming all other plasma
parameters are held constant, beyond which the fastest growing mode
has a growth rate exceeding some specified value $\gamma_{\rm min}$.
Varying a second parameter enables the construction of a stability
threshold model for each kind of unstable mode for a single
free-energy source\cite{Hellinger:2006,Verscharen:2016}. Such models
must be modified for any variation of other plasma parameters,
including minor ion densities or relative drifts between components,
which can suppress or enhance the modeled instability as well as drive
other unstable modes\cite{Price:1986,Podesta:2011b,Maruca:2012}.

These simple two-parameter models were combined with decades of
observations to demonstrate that the solar wind's evolution is bound
by long-wavelength instabilities, specifically by the mirror and CGL
firehose thresholds\cite{Kasper:2002,Matteini:2007,Bale:2009}. Chen et
al. 2016 \cite{Chen:2016} accounted for the free energy contribution
from protons, electrons, and He$^{2+}$ ($\alpha$) to long-wavelength
instability thresholds, further demonstrating that the solar wind is
well constrained by these long-wavelength instabilities and that each
plasma species contributes to the stability threshold. However, such
long-wavelength thresholds neglect instabilities arising at kinetic
scales, and in the case of the mirror mode threshold, neglect the
effects of relatively drifting components. Using these methods, the
majority of intervals were found to be stable, with only a few percent
classified as unstable.

Instead of focusing on a single free-energy source or using
long-wavelength thresholds which neglect kinetic-scale instabilities,
we identify the presence of any ion-driven instabilities using a
numerical implementation of Nyquist's instability
criterion\cite{Nyquist:1932,Klein:2017c}, which determines the number
of unstable modes supported by a specified linearized equilibrium via
a contour integral.  Of a statistically random set of Wind
observations with protons and alpha particles modeled as a collection
of drifting bi-Maxwellians, $53.7\%$ are found to be unstable.
Unstable modes preferentially arise at parallel ion-kinetic scales and
for spectra with an observed proton beam. Instabilities appear to be
pervasive in the solar wind, rather than simply serving as a boundary
that constrains its evolution, only acting on a minority of intervals.

\emph{Nyquist's Instability Criterion.}--- Nyquist's method determines
if any complex frequency solutions $[\omega,\gamma](\V{k})$ to a
dispersion relation $|D(\omega,\gamma,\V{k},\mathcal{P})|=0$ have a
positive imaginary component $\gamma > 0$ and thus are unstable for a
given wavevector $\V{k}$ and other system parameters
$\mathcal{P}$\cite{Nyquist:1932}. This is achieved by calculating the
contour integral of $|D|^{-1}$ over the upper half of the complex
frequency plane for fixed values of $\V{k}$ and $\mathcal{P}$ and
counting the number of enclosed poles via the residue theorem,
producing an integer the winding number $W_n$. If $W_n=0$, the system
is stable; if $W_n=N$, the system supports $N$ unstable modes. This
method, as well as the specific numerical implementation employed in
this work, are described in more detail in Klein et
al. 2017\cite{Klein:2017c}. This method does not report the kind of
mode driven unstable, only if an unstable mode exists. This
calculation can be performed not just to test for absolute
instability, integrating over the complex half-plane with lower
boundary $\gamma=0$, but for any minimum growth rate, performing a
contour integral with arbitrary lower boundary $\gamma=\gamma_{\rm
  min}$, yielding the number of unstable modes with growth rates
larger than $\gamma_{\rm min}$, $W_n(\V{k},\mathcal{P},\gamma_{\rm
  min})$.

To apply Nyquist's method to solar wind observations, we treat the
solar wind as a hot, magnetized plasma consisting of a collection of
drifting bi-Maxwellian populations. The linear response of this system
is described by the set of parameters $\mathcal{P}$ which includes a
normalized density $n_s/n_{\rm ref}$, drift speed relative to the
reference distribution $v_s$ normalized by the \Alfven speed
$v_A=B/\sqrt{4 \pi n_{\rm ref}m_{\rm ref}}$, parallel and
perpendicular temperatures defined by $T_{\perp s}/T_{\parallel s}$
and $T_{\parallel s}/T_{\parallel \rm ref}$, charge $q_s/q_{\rm ref}$
and mass $m_s/m_{\rm ref}$ for each component $s$, as well as a
reference plasma beta $\beta_{\parallel \rm ref}=8 \pi n_{\rm ref}
T_{\parallel \rm ref}/B^2$ and thermal speed $v_{t \rm ref}/c=\sqrt{2
  T_{\parallel \rm ref}/ m_{\rm ref} c^2}$. The linear dispersion
relation $|D|$ for such a system is calculated as a function of
wavevector $(k_\perp,k_\parallel)\rho_{\rm ref}$ normalized to the
reference gyroradius $\rho_{\rm ref}=v_{t \rm ref}/\Omega_{\rm ref}$
using the PLUME numerical solver\cite{Klein:2015a}. We calculate
$W_n(\V{k},\mathcal{P},\gamma_{\rm min})$ by numerical integration of
$|D|^{-1}$ using the proton core distribution as the reference species
and normalizing our time scales by the proton gyrofrequency $\Omega_p
= q_p B/m_p c$. For an observed $\mathcal{P}$, we calculate
$W_n(\V{k},\mathcal{P},\gamma_{\rm min})$ over a log-spaced grid
covering $(k_\perp,k_\parallel)\rho_p \in [10^{-2},10^{1}]$ and define
the unstable mode density as $\text{\dh}(\gamma_{\rm min})= \left[
  \int d\V{k} W_n(\V{k},\mathcal{P},\gamma_{\rm min})\right]/\int
d\V{k}$.


\emph{Data.}---We choose for our analysis a random set of solar wind
observations, rather than intervals associated with signatures for the
presence of instabilities\cite{Gary:2016}, selecting the first nominal
peak-tracking mode spectrum of the day measured by the Solar Wind
Experiment Faraday cup \cite{Ogilvie:1995} on the Wind spacecraft from
309 days in 2016 and 2017; data from the
magnetometer\cite{Lepping:1995,Koval:2013} is used to determine the
orientation and amplitude of the magnetic field. For each spectrum, a
nonlinear-least-squares Bi-Maxwellian fit is performed for up to three
ion components--- a proton core, proton beam, and $\alpha$
population--- using intelligent initial guesses to find the simplest
physical model that fits the data. The number of spectra with resolved
proton beams and/or an $\alpha$ population is listed in
Table~\ref{tab:occurance}. {While inclusion of electron free-energy
sources may decrease stability at fluid and kinetic scales
\cite{Chen:2016,Kunz:2018}, the details of the electron VDF will not
  significantly inhibit ion-driven instabilities. We treat the electrons as
  isotropic Maxwellians with $T_e = T_p = (2 T_{\perp p} +
  T_{\parallel p})/3$ and a drift speed necessary to ensure zero net
  current.}

\begin{table}[t]
  \begin{tabular}{ c | c | c || c | c | c | }
    & \# Spectra & \# Unstable & Mirror & CGL FH & Kinetic \\ \hline
    Total & 309 & 166 & 14 & 1 & 151 \\ \hline
    p, b, \& $\alpha$ & 189 & 130 & 12 & 0 & 118 \\
    p \& $\alpha$ & 114 & 33 & 2 & 1 & 30 \\
    p \& b & 5 & 3 & 0 & 0 & 3 \\
    p & 1 & 0 & 0 & 0 & 0 \\
    \hline
  \end{tabular}
\caption{\label{tab:occurance}Total number, and number of unstable,
  spectra. The results are divided between cases with and without
  resolved proton beam and/or $\alpha$ components. The unstable
  spectra are further divided into mirror, CGL firehose, and
  ion-kinetic-scale instabilities.}
\end{table}

\begin{figure}[t]
\resizebox{4.1in}{!}
{\includegraphics*[0.5in,0.4in][4.5in,2.88in]{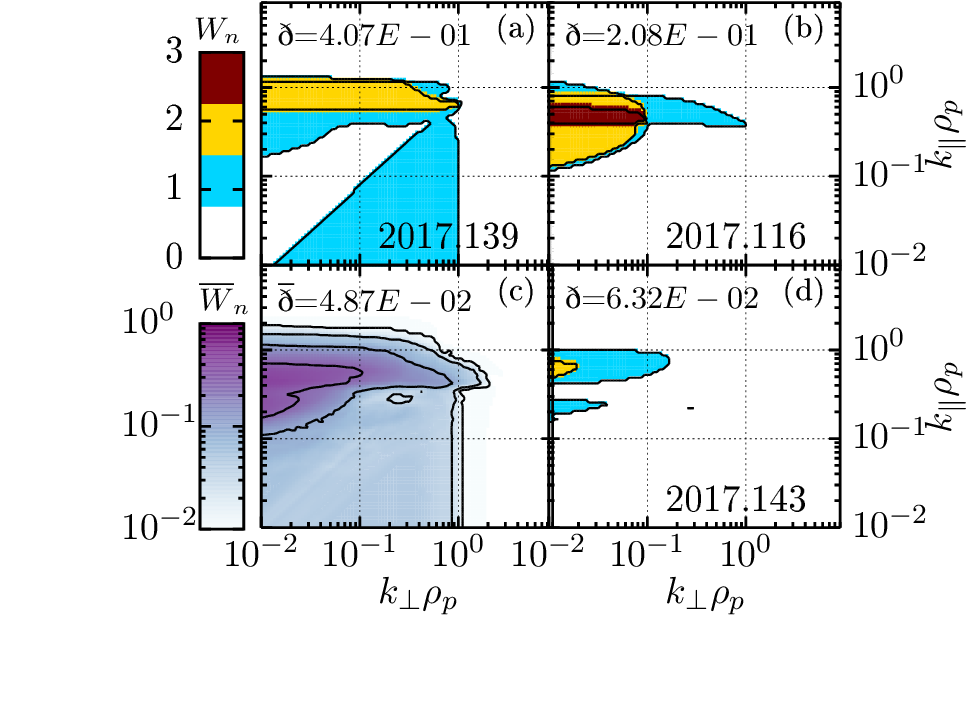}}
\caption{ \label{fig:example} (a,b,d) The number of unstable modes
  with $\gamma_{\rm min}>0$ as a function of wavevector $\V{k}\rho_p$
  for three example spectra. (c) The mean winding number averaged over
  the 309 observed spectra.}
\end{figure}

For spectra without a proton beam population, values for 7
dimensionless parameters are extracted from Bi-Maxwellian fits:
$\beta_{\parallel p}$, $v_{t p}/c$, $T_{\perp p}/T_{\parallel p}$,
$T_{\perp \alpha}/T_{\parallel \alpha}$, $T_{\parallel
  \alpha}/T_{\parallel p}$, $n_\alpha/n_p$, and $v_\alpha/v_A$.  For
spectra with a proton beam, 4 additional parameters are used:
$T_{\perp b}/T_{\parallel b}$, $T_{\parallel b}/T_{\parallel p}$,
$n_b/n_p$, and $v_b/v_A$.  {Mean values of these
  parameters, given in Table~\ref{tab:param}, are consistent with
  previous statistical studies of solar wind
  observations\cite{Wilson:2018}, though the inclusion of proton beams
  in this work reduces $T_{\parallel p}$ compared to studies which
  assume a single proton population.}  We calculate
$W_n(\V{k}\rho_p,\mathcal{P},\gamma_{\rm min}=0)$ as a function of
$(k_\perp,k_\parallel)\rho_p$; example winding number distributions
and unstable mode densities \dh~for three unstable spectra are shown
in Fig.~\ref{fig:example}, as well as the mean winding number
$\overline{W}_n(\V{k}\rho_p,\gamma_{\rm min}=0)$ averaged over all 309
spectra.

\begin{table*}[t]
\begin{tabular}{ c || c | c || c | c | c || c | c|| c | c || c | c|}
& $\beta_{\parallel p}$ & $10^{4} v_{t p}/c$ & $T_{\perp p}/T_{\parallel p}$ & $T_{\perp \alpha}/T_{\parallel \alpha}$ & $T_{\perp b}/T_{\parallel b}$ &$T_{\parallel \alpha}/T_{\parallel p}$ & $T_{\parallel b}/T_{\parallel p}$ &$n_\alpha/n_p$ & $n_b/n_p$ & $|v_\alpha|/v_A$ & $|v_b|/v_A$   \\ \hline
Total &  0.60&  1.07&  1.57&  0.96&  1.48& 10.89&  2.72&  0.04&  0.43&  0.31&  0.84\\ \hline
Stable &  0.50&  0.91&  1.12&  1.03&  1.39&  5.24&  2.35&  0.04&  0.41&  0.16&  0.73\\ \hline
Unstable &  0.68&  1.21&  1.96&  0.90&  1.52& 15.74&  2.88&  0.05&  0.44&  0.44&  0.89\\ \hline \hline
$\Delta X_{p,\alpha,b} (\%)$ & 19.12& 13.46& 50.59&-21.06&  8.45& 64.27& 20.83&  2.61&  2.90& 61.57& 21.84\\ \hline
$\Delta X_{p,\alpha} (\%)$ &132.53& 57.59&-26.77& 14.16& --- & 26.46& --- & 18.10& --- & 77.44& --- \\ \hline
\end{tabular}
\caption{\label{tab:param} Mean plasma parameters for the 309 observed
  spectra (top row), for the stable and unstable spectra (second and
  third), and the normalized difference of the parameters $\Delta X$
  between stable and unstable spectra (fourth and fifth).}
\end{table*}

\begin{figure}[t]
\resizebox{3.5in}{!}  {\includegraphics*[0.4in,0.2in][3.4in,2.1in]
  {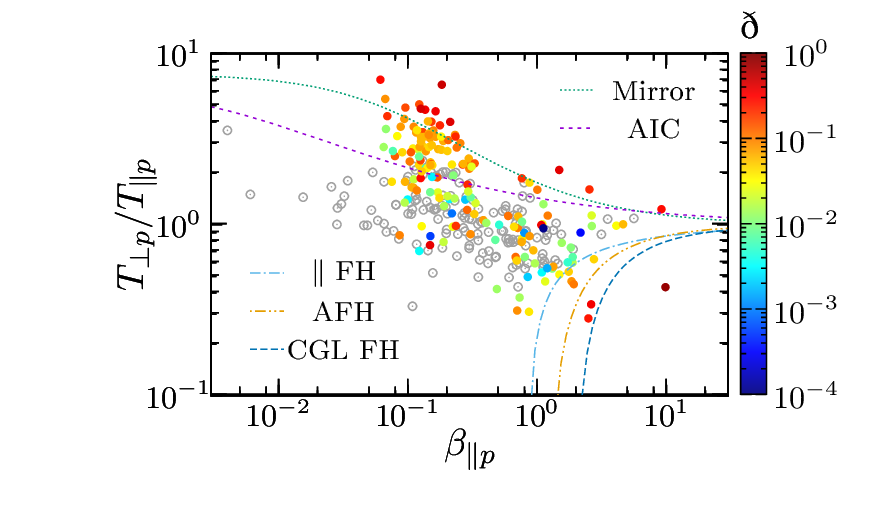}}
\caption{ \label{fig:beta_ani_dist} The ($\beta_{\parallel p},T_{\perp
    p}/T_{\parallel p}$) distribution of the observed spectra; color
  indicates the unstable mode density \dh, and grey indicates a stable
  spectrum.}
\end{figure}

\emph{Occurrence of Instability}--- We find that $53.7 \%$ of the
randomly selected spectra have \dh$(\gamma_{\rm min}=0)>0$ and thus
support at least one growing mode in $(k_\perp,k_\parallel)\rho_p \in
[10^{-2},10^{1}]$. Considering the spectra with (without) a proton
beam, $70.0\%$ ($28.7\%$) are unstable; a summary of the number of
unstable modes as a function of the resolved components is presented
in Table~\ref{tab:occurance}. Fig.~\ref{fig:beta_ani_dist} illustrates
the ($\beta_{\parallel p},T_{\perp p}/T_{\parallel p}$) distribution
of the 309 spectra; unstable spectra are color-coded by the associated
unstable mode density \dh~and stable spectra are plotted in grey. The
stability thresholds for proton-temperature anisotropy-driven
instabilities with $\gamma_{\rm
  min}=10^{-3}\Omega_p$\cite{Verscharen:2016} are included for
context.

The mean winding number $\overline{W}_n(\V{k}\rho_p,0)$,
Fig.~\ref{fig:example}(c), shows that most unstable modes arise at
parallel wavevectors near ion kinetic scales ($k_\perp \rho_p <
k_\parallel \rho_p \lesssim 1$) though there exist a finite number of
unstable modes at long wavelengths and/or at more oblique
wavevectors. The abrupt cutoff of $\overline{W}_n$ beyond
$k\rho_p\approx 1$ is due to our model's lack of electron free-energy
sources, which are necessary to drive instabilities between ion
and electron kinetic scales.

\begin{figure*}[t]
\resizebox{6.in}{!}
{\includegraphics*[0.375in,0.01in][5.55in,2.05in]
{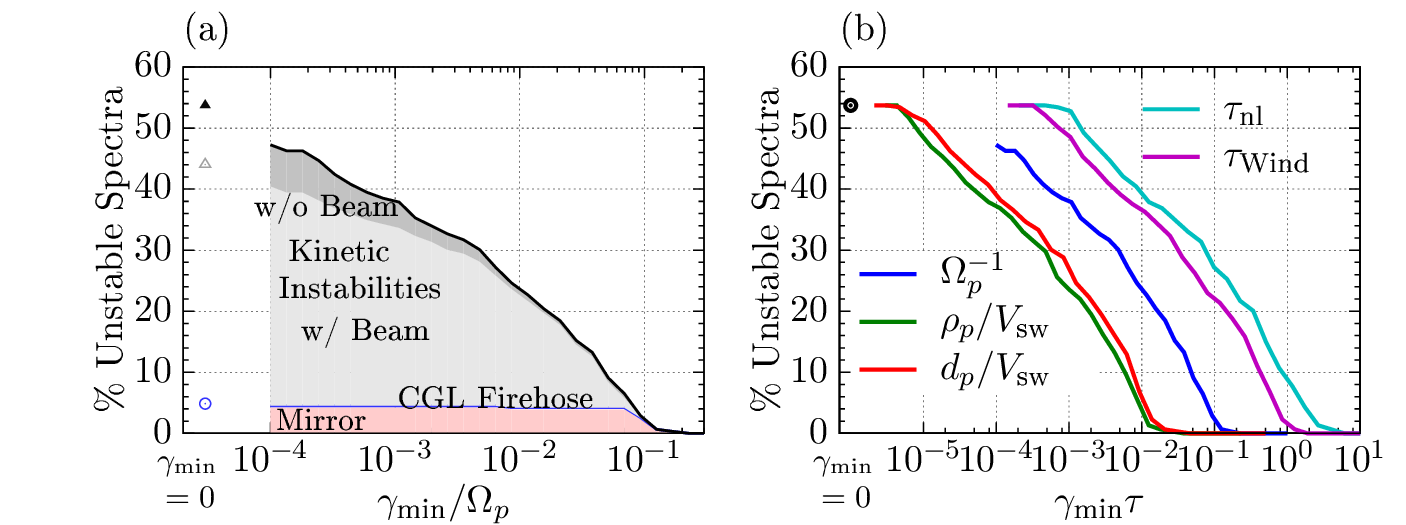}}
\caption{ \label{fig:distribution} The fraction of observed spectra
  supporting unstable modes with growth rate exceeding
  $\gamma_{\textrm{min}}$. (a) The spectra are divided
  according to the instability classification presented in the
  text. (b) The minimum growth rate distribution is re-scaled
  by selected timescales $\tau$.}
\end{figure*}

To determine what kinds of instabilities arise for a given spectrum,
we inspect $W_n(\V{k}\rho_p,\mathcal{P},0)$ for the 166 unstable
spectra. For the mirror instability, the long-wavelength
threshold\cite{Hellinger:2007} can not be simply applied, as it does
not account for the effects of relative drift between
distributions. Instead, we identified 14 spectra that have unstable
modes with $|k \rho_p|$ extending from long-wavelengths up to the
proton gyroscale covering oblique angles, $k_\perp >
k_\parallel$. These intervals are classified as mirror unstable; an
example of such a spectrum is found in Fig.~\ref{fig:example}(a). For
each mirror unstable case, there also exist kinetic instabilities with
$k_\parallel \rho_p \lesssim 1$, in agreement with the canonical
$T_{\perp p}/T_{\parallel p}>1$ mirror unstable distribution, (e.g.
Fig. 2(c) of Klein et al. 2017\cite{Klein:2017c}). One spectrum, not
shown, exceeds the long-wavelength CGL firehose
threshold\cite{Kunz:2015} and has a winding number distribution
similar to the canonical case (e.g. Fig. 2(f) of Klein et al. 2017),
driving unstable modes for nearly all wavevectors with $k \rho_p<1$,
one (two) mode(s) for $k_\perp >(<) k_\parallel$. We classify the
remaining 151 unstable spectra with growing modes satisfying $k_\perp
\rho_p< k_\parallel \rho_p\lesssim 1$ as kinetic; two example $W_n$
distributions for these kinetic cases are shown in
Fig.~\ref{fig:example}(b) and (d). The instability classification as a
function of resolved ion components is given in
Table~\ref{tab:occurance}.

Using this classification scheme, we repeat our analysis for a range
of minimum growth rates $\gamma_{\rm min} \in [10^{-4},10^0]\Omega_p$,
shown in Fig.~\ref{fig:distribution}. We see that (black line in
Fig.~\ref{fig:distribution}(a)) the fraction of unstable spectra
decreases with an increase in $\gamma_{\rm min}$, with no spectrum
having growth rates exceeding $\gamma>0.2 \Omega_p$. The number of
mirror and CGL firehose unstable modes (red and blue regions) remains
constant with increasing $\gamma_{\rm min}$ up to $0.1 \Omega_p$. Most
of the kinetic instabilities associated with spectra without proton
beams (dark grey) are limited to growth rates less than $10^{-2}
\Omega_p$, while a decreasing fraction of the unstable spectra with
proton beams (light grey) persists to $0.1 \Omega_p$.

\emph{Instability Timescales}--- To compare $\gamma_{\rm min}$ with
timescales other than $\Omega_p^{-1}$, we calculate the fraction of
unstable spectra as a function of four additional time scales; the
advected proton gyroscale timescale $\rho_p/v_{\rm SW}$, the advected
proton inertial length timescale $d_p/v_{\rm SW}=v_A/( \Omega_p v_{\rm
  SW})$, the Faraday cup measurement period $\tau_{\rm Wind} =
92~\text{s}$, and $\tau_{\rm nl}= (k_0 \rho_p)^{-1/3} \rho_p/v_A$, an
estimate for the nonlinear turbulent energy transfer time at the
proton gyroscale $k_\perp \rho_p = 1$ assuming a critically balanced
cascade of energy\cite{Goldreich:1995,Mallet:2015} from an outer scale
$k_0 \rho_p = 10^{-4}$. For each spectrum, the values for these
timescales are calculated and the unstable mode density
\dh$(\gamma_{\rm min} /\Omega_p)$ is interpolated onto a log-spaced
grid for \dh$(\gamma_{\rm min} \tau)$.  This distribution is averaged
over the 309 spectra to calculate the fraction of unstable spectra as
a function of $\tau$, shown in Fig.~\ref{fig:distribution}(b).

The unstable modes typically have growth rates slower than ion-kinetic
timescales. Nearly all unstable spectra have growth rates slower than
a hundredth of $\rho_p/v_{\rm SW}$ or $d_p/v_{\rm SW}$,
{indicating any growing ion-kinetic-scale structure
  associated with instabilities will be static in the spacecraft
  frame}. As nearly all unstable spectra have growth rates slower than
$92 {\rm s}$ the nominal spectra selected for this work are in steady
state with respect to any instability induced evolution. Less than
$10\%$ of the spectra have growth rates faster than $\tau_{\rm nl}$,
indicating that only a small fraction of the instabilities act quickly
enough to compete with ion-scale damping processes.

{\emph{Parametric Dependence.}}---We wish to determine any
relation between a velocity distribution's bulk parameters and its
stability. Given the high-dimensionality of the parameter space --- $3
+ 4 (N_{\rm ion} - 1)$ values for $N_{\rm ion}$ resolved ion
components--- it is difficult to determine the relative importance of
a given parameter; previous attempts typically focused on the effects
of a handful of parameters, e.g. $\beta_{\parallel p}$ and
$T_{\perp,p}/T_{\parallel,p}$. To ascertain any relation, we calculate
the normalized difference
\begin{equation}
\Delta X \equiv \frac{\bar{X}_{\rm unstable} -
  \bar{X}_{\rm stable}}{ \bar{X}_{\rm total}}
\end{equation}
with $X$ drawn from the ion bulk parameters; $\bar{X}_{\rm total}$,
$\bar{X}_{\rm unstable}$ and $\bar{X}_{\rm stable}$ are the mean value
of $X$ averaged over all spectra, over the unstable spectra, and over
the stable spectra, with stability determined using $\gamma_{\rm
  min}=0$. Selection of larger $\gamma_{\rm min}/\Omega_p$ does not
qualitatively alter these results. We calculate $\Delta X$ using two
disjoint sub-sets of data; spectra with a resolved alpha distribution
and proton core, or spectra with all three ion components resolved.
Values of $\Delta X$ are presented in Table~\ref{tab:param}.

Unstable spectra both without and with proton beams have higher mean
alpha drift velocities $v_\alpha/v_A$ than stable spectra, indicating
that the free energy associated with the larger relative drift between
the protons and alphas is important in driving instabilities. The mean
core proton temperature anisotropy $T_{\perp p}/T_{\parallel p}$ for
unstable spectra is significantly decreased (increased) from isotropy
for cases without (with) a proton beam.  This reduction of temperature
anisotropy is potentially due to the beam having relaxed into the
proton core, leading to an increased $T_{\parallel p}$.

For the no-proton-beam case, $\beta_{\parallel p}$ is significantly
larger for the unstable spectra, with a $132\%$ increase compared to
stable spectra. The normalized core proton thermal speed $v_{tp}/c$,
our dimensionless proxy for the {parallel} core proton
temperature, is also significantly larger. Combined with the
normalized difference $\Delta |v_\alpha|/v_A$, this indicates that
parallel free energy is {important} for driving these
systems unstable.

For spectra with proton beams, $T_{\parallel \alpha}/T_{\parallel p}$
is increased for unstable spectra. The proton beam is also slightly
hotter while the alpha temperature anisotropy $T_{\perp
  \alpha}/T_{\parallel \alpha}$ is slightly decreased. The values of
the other proton beam parameters are only marginally increased for
unstable spectra.

{\emph{Effects of Uncertainty}}--- To consider the
robustness of this method against measurement uncertainty, we follow
Klein et al. 2017\cite{Klein:2017c} and repeat our instability
analysis on an ensemble of 100 Monte Carlo variations of $\mathcal{P}$
for each of the 309 observed spectra. Each observed dimensional
quantity from which $\mathcal{P}$ is composed is replaced by a
Gaussian-distributed random variable with a mean of the original
quantity and a standard deviation of $10\%$.  The width of the random
variable distribution is motivated by measurement uncertainties
found for instance by Kasper et al. 2006\cite{Kasper:2006}.  For these 31,209 values of
$\mathcal{P}$, {$56.0\%$} are unstable, qualitatively
similar to $53.7\%$ calculated from the observed spectra. For the
ensembles corresponding to stable observations,
$\mathcal{P}(\textrm{\dh}_0=0)$, an average of {$83.6\%$}
of the elements are stable; for $\mathcal{P}(\textrm{\dh}_0\neq 0)$,
an average of {$90.5\%$} are unstable. Of the
\dh$_0\neq0$ ensembles, $0.6\%$ have a majority of their elements stable, while
   {$8.3\%$} of the \dh$_0=0$ ensembles are majority
   unstable.



In addition to measurement uncertainty, our results may be affected by
unresolved proton beams with small $n_b/n_p$ or $v_b/v_A$, or by the
assumption of bi-Maxwellian
distributions\cite{Dum:1980,Isenberg:2012}. Repeating this work with a
dispersion relation which neglects analytic forms and captures
non-Maxwellian features\cite{Verscharen:2018} will enable more
accurate determination of solar wind stability.



{ \emph{Conclusions.}}---We assess the stability of 309
randomly selected solar wind spectra with ion components modeled as a
collection of drifting bi-Maxwellians using Nyquist's instability
criterion and find $53.7\%$ are unstable. This mode-agnostic method
includes the effects of ion drifts and temperature anisotropies,
contrasting with previously employed threshold models that identify
only a small fraction of solar wind intervals as unstable.  This
method identifies the same instabilities as traditional Vlasov
studies, but does not require a priori knowledge of which linear modes
are unstable, allowing for automated analysis.  The unstable modes
identified using Nyquist's criterion are primarily kinetic, with
$k_\perp \rho_p < k_\parallel \rho_p \lesssim 1$; only $4.5\%$ of
the observed spectra have long-wavelength instabilities. The maximum
growth rate for these unstable modes is slower than measurement and
ion-kinetic timescales. The mean alpha drift speed for unstable
spectra is larger than for stable spectra, and the ratio $T_{\perp
  p}/T_{\parallel p}$ for unstable spectra is further from isotropy.
The majority of the unstable spectra have a resolved proton beam
component.

Further study is needed to assess the effects of this profusion of
instabilities. While a majority of observed spectra are unstable, it
remains unclear from this initial study if all the inferred
instabilities are dynamically important, or simply a biproduct of
other processes. The resonant instabilities which comprise the
majority of the unstable spectra do not act as efficiently as
long-wavelength instabilities to return the plasma toward isotropy and
therefore may not constrain the dynamics of the solar wind's
evolution. This may be an effect of slower growth rates, smaller
regions of wavevector space being driven unstable, or departures from
the assumed bi-Maxwellian distribution affecting resonance conditions.

One way to discern if these instabilities are are continuously
generated or a remnant of processes in the near-Sun environment, and
how their role in solar wind dynamics changes at varying distances
from the Sun, will be to combine this automated instability detection
method with forthcoming measurements from Parker Solar
Probe\cite{Fox:2016} and Solar Orbiter\cite{Muller:2013}.

\emph{Acknowledgments.}---
The spectrum data used in this project were taken from the
\texttt{WI\_SW-ION-DIST\_SWE-FARADAY} database at
\url{cdaweb.gsfc.nasa.gov}. 
The authors would like to
acknowledge inspiring conversations with Alfred Mallet. This work was
supported by NASA grant NNX14AR78G, with K.G. Klein receiving
additional support from NASA grant NNX16AG81G.


\begin{thebibliography}{38}%
\makeatletter
\providecommand \@ifxundefined [1]{%
 \@ifx{#1\undefined}
}%
\providecommand \@ifnum [1]{%
 \ifnum #1\expandafter \@firstoftwo
 \else \expandafter \@secondoftwo
 \fi
}%
\providecommand \@ifx [1]{%
 \ifx #1\expandafter \@firstoftwo
 \else \expandafter \@secondoftwo
 \fi
}%
\providecommand \natexlab [1]{#1}%
\providecommand \enquote  [1]{``#1''}%
\providecommand \bibnamefont  [1]{#1}%
\providecommand \bibfnamefont [1]{#1}%
\providecommand \citenamefont [1]{#1}%
\providecommand \href@noop [0]{\@secondoftwo}%
\providecommand \href [0]{\begingroup \@sanitize@url \@href}%
\providecommand \@href[1]{\@@startlink{#1}\@@href}%
\providecommand \@@href[1]{\endgroup#1\@@endlink}%
\providecommand \@sanitize@url [0]{\catcode `\\12\catcode `\$12\catcode
  `\&12\catcode `\#12\catcode `\^12\catcode `\_12\catcode `\%12\relax}%
\providecommand \@@startlink[1]{}%
\providecommand \@@endlink[0]{}%
\providecommand \url  [0]{\begingroup\@sanitize@url \@url }%
\providecommand \@url [1]{\endgroup\@href {#1}{\urlprefix }}%
\providecommand \urlprefix  [0]{URL }%
\providecommand \Eprint [0]{\href }%
\providecommand \doibase [0]{http://dx.doi.org/}%
\providecommand \selectlanguage [0]{\@gobble}%
\providecommand \bibinfo  [0]{\@secondoftwo}%
\providecommand \bibfield  [0]{\@secondoftwo}%
\providecommand \translation [1]{[#1]}%
\providecommand \BibitemOpen [0]{}%
\providecommand \bibitemStop [0]{}%
\providecommand \bibitemNoStop [0]{.\EOS\space}%
\providecommand \EOS [0]{\spacefactor3000\relax}%
\providecommand \BibitemShut  [1]{\csname bibitem#1\endcsname}%
\let\auto@bib@innerbib\@empty
\bibitem [{\citenamefont {{Kennel}}\ and\ \citenamefont
  {{Petschek}}(1966)}]{Kennel:1966}%
  \BibitemOpen
  \bibfield  {author} {\bibinfo {author} {\bibfnamefont {C.~F.}\ \bibnamefont
  {{Kennel}}}\ and\ \bibinfo {author} {\bibfnamefont {H.~E.}\ \bibnamefont
  {{Petschek}}},\ }\href@noop {} {\bibfield  {journal} {\bibinfo  {journal}
  {J.~Geophys.~Res.}\ }\textbf {\bibinfo {volume} {71}},\ \bibinfo {pages} {1}
  (\bibinfo {year} {1966})}\BibitemShut {NoStop}%
\bibitem [{\citenamefont {{Davidson}}\ and\ \citenamefont
  {{Ogden}}(1975)}]{Davidson:1975}%
  \BibitemOpen
  \bibfield  {author} {\bibinfo {author} {\bibfnamefont {R.~C.}\ \bibnamefont
  {{Davidson}}}\ and\ \bibinfo {author} {\bibfnamefont {J.~M.}\ \bibnamefont
  {{Ogden}}},\ }\href {\doibase 10.1063/1.861253} {\bibfield  {journal}
  {\bibinfo  {journal} {Physics of Fluids}\ }\textbf {\bibinfo {volume} {18}},\
  \bibinfo {pages} {1045} (\bibinfo {year} {1975})}\BibitemShut {NoStop}%
\bibitem [{\citenamefont {{Tajiri}}(1967)}]{Tajiri:1967}%
  \BibitemOpen
  \bibfield  {author} {\bibinfo {author} {\bibfnamefont {M.}~\bibnamefont
  {{Tajiri}}},\ }\href@noop {} {\bibfield  {journal} {\bibinfo  {journal}
  {Journal of the Physical Society of Japan}\ }\textbf {\bibinfo {volume}
  {22}},\ \bibinfo {pages} {1482} (\bibinfo {year} {1967})}\BibitemShut
  {NoStop}%
\bibitem [{\citenamefont {{Southwood}}\ and\ \citenamefont
  {{Kivelson}}(1993)}]{Southwood:1993}%
  \BibitemOpen
  \bibfield  {author} {\bibinfo {author} {\bibfnamefont {D.~J.}\ \bibnamefont
  {{Southwood}}}\ and\ \bibinfo {author} {\bibfnamefont {M.~G.}\ \bibnamefont
  {{Kivelson}}},\ }\href {\doibase 10.1029/92JA02837} {\bibfield  {journal}
  {\bibinfo  {journal} {J.~Geophys.~Res.}\ }\textbf {\bibinfo {volume} {98}},\
  \bibinfo {pages} {9181} (\bibinfo {year} {1993})}\BibitemShut {NoStop}%
\bibitem [{\citenamefont {{Kivelson}}\ and\ \citenamefont
  {{Southwood}}(1996)}]{Kivelson:1996}%
  \BibitemOpen
  \bibfield  {author} {\bibinfo {author} {\bibfnamefont {M.~G.}\ \bibnamefont
  {{Kivelson}}}\ and\ \bibinfo {author} {\bibfnamefont {D.~J.}\ \bibnamefont
  {{Southwood}}},\ }\href {\doibase 10.1029/96JA01407} {\bibfield  {journal}
  {\bibinfo  {journal} {J.~Geophys.~Res.}\ }\textbf {\bibinfo {volume} {101}},\
  \bibinfo {pages} {17365} (\bibinfo {year} {1996})}\BibitemShut {NoStop}%
\bibitem [{\citenamefont {{Quest}}\ and\ \citenamefont
  {{Shapiro}}(1996)}]{Quest:1996}%
  \BibitemOpen
  \bibfield  {author} {\bibinfo {author} {\bibfnamefont {K.~B.}\ \bibnamefont
  {{Quest}}}\ and\ \bibinfo {author} {\bibfnamefont {V.~D.}\ \bibnamefont
  {{Shapiro}}},\ }\href {\doibase 10.1029/96JA01534} {\bibfield  {journal}
  {\bibinfo  {journal} {J.~Geophys.~Res.}\ }\textbf {\bibinfo {volume} {101}},\
  \bibinfo {pages} {24457} (\bibinfo {year} {1996})}\BibitemShut {NoStop}%
\bibitem [{\citenamefont {{Gary}}\ \emph {et~al.}(1998)\citenamefont {{Gary}},
  \citenamefont {{Li}}, \citenamefont {{O'Rourke}},\ and\ \citenamefont
  {{Winske}}}]{Gary:1998}%
  \BibitemOpen
  \bibfield  {author} {\bibinfo {author} {\bibfnamefont {S.~P.}\ \bibnamefont
  {{Gary}}}, \bibinfo {author} {\bibfnamefont {H.}~\bibnamefont {{Li}}},
  \bibinfo {author} {\bibfnamefont {S.}~\bibnamefont {{O'Rourke}}}, \ and\
  \bibinfo {author} {\bibfnamefont {D.}~\bibnamefont {{Winske}}},\ }\href
  {\doibase 10.1029/98JA01174} {\bibfield  {journal} {\bibinfo  {journal}
  {J.~Geophys.~Res.}\ }\textbf {\bibinfo {volume} {103}},\ \bibinfo {pages}
  {14567} (\bibinfo {year} {1998})}\BibitemShut {NoStop}%
\bibitem [{\citenamefont {{Hellinger}}\ and\ \citenamefont
  {{Matsumoto}}(2000)}]{Hellinger:2000}%
  \BibitemOpen
  \bibfield  {author} {\bibinfo {author} {\bibfnamefont {P.}~\bibnamefont
  {{Hellinger}}}\ and\ \bibinfo {author} {\bibfnamefont {H.}~\bibnamefont
  {{Matsumoto}}},\ }\href {\doibase 10.1029/1999JA000297} {\bibfield  {journal}
  {\bibinfo  {journal} {J.~Geophys.~Res.}\ }\textbf {\bibinfo {volume} {105}},\
  \bibinfo {pages} {10519} (\bibinfo {year} {2000})}\BibitemShut {NoStop}%
\bibitem [{\citenamefont {{Chew}}\ \emph {et~al.}(1956)\citenamefont {{Chew}},
  \citenamefont {{Goldberger}},\ and\ \citenamefont {{Low}}}]{Chew:1956}%
  \BibitemOpen
  \bibfield  {author} {\bibinfo {author} {\bibfnamefont {G.~F.}\ \bibnamefont
  {{Chew}}}, \bibinfo {author} {\bibfnamefont {M.~L.}\ \bibnamefont
  {{Goldberger}}}, \ and\ \bibinfo {author} {\bibfnamefont {F.~E.}\
  \bibnamefont {{Low}}},\ }\href {\doibase 10.1098/rspa.1956.0116} {\bibfield
  {journal} {\bibinfo  {journal} {Royal Society of London Proceedings Series
  A}\ }\textbf {\bibinfo {volume} {236}},\ \bibinfo {pages} {112} (\bibinfo
  {year} {1956})}\BibitemShut {NoStop}%
\bibitem [{\citenamefont {Yoon}(2017)}]{Yoon:2017}%
  \BibitemOpen
  \bibfield  {author} {\bibinfo {author} {\bibfnamefont {P.~H.}\ \bibnamefont
  {Yoon}},\ }\href {\doibase 10.1007/s41614-017-0006-1} {\bibfield  {journal}
  {\bibinfo  {journal} {Reviews of Modern Plasma Physics}\ }\textbf {\bibinfo
  {volume} {1}},\ \bibinfo {pages} {4} (\bibinfo {year} {2017})}\BibitemShut
  {NoStop}%
\bibitem [{\citenamefont {{Hellinger}}\ \emph {et~al.}(2006)\citenamefont
  {{Hellinger}}, \citenamefont {{Tr{\'a}vn{\'{\i}}{\v c}ek}}, \citenamefont
  {{Kasper}},\ and\ \citenamefont {{Lazarus}}}]{Hellinger:2006}%
  \BibitemOpen
  \bibfield  {author} {\bibinfo {author} {\bibfnamefont {P.}~\bibnamefont
  {{Hellinger}}}, \bibinfo {author} {\bibfnamefont {P.}~\bibnamefont
  {{Tr{\'a}vn{\'{\i}}{\v c}ek}}}, \bibinfo {author} {\bibfnamefont {J.~C.}\
  \bibnamefont {{Kasper}}}, \ and\ \bibinfo {author} {\bibfnamefont {A.~J.}\
  \bibnamefont {{Lazarus}}},\ }\href {\doibase 10.1029/2006GL025925} {\bibfield
   {journal} {\bibinfo  {journal} {Geophys.~Res.~Lett.}\ }\textbf {\bibinfo
  {volume} {33}},\ \bibinfo {eid} {L09101} (\bibinfo {year}
  {2006})}\BibitemShut {NoStop}%
\bibitem [{\citenamefont {{Verscharen}}\ \emph {et~al.}(2016)\citenamefont
  {{Verscharen}}, \citenamefont {{Chandran}}, \citenamefont {{Klein}},\ and\
  \citenamefont {{Quataert}}}]{Verscharen:2016}%
  \BibitemOpen
  \bibfield  {author} {\bibinfo {author} {\bibfnamefont {D.}~\bibnamefont
  {{Verscharen}}}, \bibinfo {author} {\bibfnamefont {B.~D.~G.}\ \bibnamefont
  {{Chandran}}}, \bibinfo {author} {\bibfnamefont {K.~G.}\ \bibnamefont
  {{Klein}}}, \ and\ \bibinfo {author} {\bibfnamefont {E.}~\bibnamefont
  {{Quataert}}},\ }\href {\doibase 10.3847/0004-637X/831/2/128} {\bibfield
  {journal} {\bibinfo  {journal} {Astrophys.~J.}\ }\textbf {\bibinfo {volume}
  {831}},\ \bibinfo {eid} {128} (\bibinfo {year} {2016})},\ \Eprint
  {http://arxiv.org/abs/1605.07143} {arXiv:1605.07143 [physics.space-ph]}
  \BibitemShut {NoStop}%
\bibitem [{\citenamefont {{Price}}\ \emph {et~al.}(1986)\citenamefont
  {{Price}}, \citenamefont {{Swift}},\ and\ \citenamefont
  {{Lee}}}]{Price:1986}%
  \BibitemOpen
  \bibfield  {author} {\bibinfo {author} {\bibfnamefont {C.~P.}\ \bibnamefont
  {{Price}}}, \bibinfo {author} {\bibfnamefont {D.~W.}\ \bibnamefont
  {{Swift}}}, \ and\ \bibinfo {author} {\bibfnamefont {L.-C.}\ \bibnamefont
  {{Lee}}},\ }\href {\doibase 10.1029/JA091iA01p00101} {\bibfield  {journal}
  {\bibinfo  {journal} {J.~Geophys.~Res.}\ }\textbf {\bibinfo {volume} {91}},\
  \bibinfo {pages} {101} (\bibinfo {year} {1986})}\BibitemShut {NoStop}%
\bibitem [{\citenamefont {{Podesta}}\ and\ \citenamefont
  {{Gary}}(2011)}]{Podesta:2011b}%
  \BibitemOpen
  \bibfield  {author} {\bibinfo {author} {\bibfnamefont {J.~J.}\ \bibnamefont
  {{Podesta}}}\ and\ \bibinfo {author} {\bibfnamefont {S.~P.}\ \bibnamefont
  {{Gary}}},\ }\href {\doibase 10.1088/0004-637X/742/1/41} {\bibfield
  {journal} {\bibinfo  {journal} {Astrophys.~J.}\ }\textbf {\bibinfo {volume}
  {742}},\ \bibinfo {eid} {41} (\bibinfo {year} {2011})}\BibitemShut {NoStop}%
\bibitem [{\citenamefont {{Maruca}}\ \emph {et~al.}(2012)\citenamefont
  {{Maruca}}, \citenamefont {{Kasper}},\ and\ \citenamefont
  {{Gary}}}]{Maruca:2012}%
  \BibitemOpen
  \bibfield  {author} {\bibinfo {author} {\bibfnamefont {B.~A.}\ \bibnamefont
  {{Maruca}}}, \bibinfo {author} {\bibfnamefont {J.~C.}\ \bibnamefont
  {{Kasper}}}, \ and\ \bibinfo {author} {\bibfnamefont {S.~P.}\ \bibnamefont
  {{Gary}}},\ }\href {\doibase 10.1088/0004-637X/748/2/137} {\bibfield
  {journal} {\bibinfo  {journal} {Astrophys.~J.}\ }\textbf {\bibinfo {volume}
  {748}},\ \bibinfo {eid} {137} (\bibinfo {year} {2012})}\BibitemShut {NoStop}%
\bibitem [{\citenamefont {{Kasper}}\ \emph {et~al.}(2002)\citenamefont
  {{Kasper}}, \citenamefont {{Lazarus}},\ and\ \citenamefont
  {{Gary}}}]{Kasper:2002}%
  \BibitemOpen
  \bibfield  {author} {\bibinfo {author} {\bibfnamefont {J.~C.}\ \bibnamefont
  {{Kasper}}}, \bibinfo {author} {\bibfnamefont {A.~J.}\ \bibnamefont
  {{Lazarus}}}, \ and\ \bibinfo {author} {\bibfnamefont {S.~P.}\ \bibnamefont
  {{Gary}}},\ }\href {\doibase 10.1029/2002GL015128} {\bibfield  {journal}
  {\bibinfo  {journal} {Geophys.~Res.~Lett.}\ }\textbf {\bibinfo {volume}
  {29}},\ \bibinfo {eid} {1839} (\bibinfo {year} {2002})}\BibitemShut {NoStop}%
\bibitem [{\citenamefont {{Matteini}}\ \emph {et~al.}(2007)\citenamefont
  {{Matteini}}, \citenamefont {{Landi}}, \citenamefont {{Hellinger}},
  \citenamefont {{Pantellini}}, \citenamefont {{Maksimovic}}, \citenamefont
  {{Velli}}, \citenamefont {{Goldstein}},\ and\ \citenamefont
  {{Marsch}}}]{Matteini:2007}%
  \BibitemOpen
  \bibfield  {author} {\bibinfo {author} {\bibfnamefont {L.}~\bibnamefont
  {{Matteini}}}, \bibinfo {author} {\bibfnamefont {S.}~\bibnamefont {{Landi}}},
  \bibinfo {author} {\bibfnamefont {P.}~\bibnamefont {{Hellinger}}}, \bibinfo
  {author} {\bibfnamefont {F.}~\bibnamefont {{Pantellini}}}, \bibinfo {author}
  {\bibfnamefont {M.}~\bibnamefont {{Maksimovic}}}, \bibinfo {author}
  {\bibfnamefont {M.}~\bibnamefont {{Velli}}}, \bibinfo {author} {\bibfnamefont
  {B.~E.}\ \bibnamefont {{Goldstein}}}, \ and\ \bibinfo {author} {\bibfnamefont
  {E.}~\bibnamefont {{Marsch}}},\ }\href {\doibase 10.1029/2007GL030920}
  {\bibfield  {journal} {\bibinfo  {journal} {Geophys.~Res.~Lett.}\ }\textbf
  {\bibinfo {volume} {34}},\ \bibinfo {eid} {L20105} (\bibinfo {year}
  {2007})}\BibitemShut {NoStop}%
\bibitem [{\citenamefont {{Bale}}\ \emph {et~al.}(2009)\citenamefont {{Bale}},
  \citenamefont {{Kasper}}, \citenamefont {{Howes}}, \citenamefont
  {{Quataert}}, \citenamefont {{Salem}},\ and\ \citenamefont
  {{Sundkvist}}}]{Bale:2009}%
  \BibitemOpen
  \bibfield  {author} {\bibinfo {author} {\bibfnamefont {S.~D.}\ \bibnamefont
  {{Bale}}}, \bibinfo {author} {\bibfnamefont {J.~C.}\ \bibnamefont
  {{Kasper}}}, \bibinfo {author} {\bibfnamefont {G.~G.}\ \bibnamefont
  {{Howes}}}, \bibinfo {author} {\bibfnamefont {E.}~\bibnamefont {{Quataert}}},
  \bibinfo {author} {\bibfnamefont {C.}~\bibnamefont {{Salem}}}, \ and\
  \bibinfo {author} {\bibfnamefont {D.}~\bibnamefont {{Sundkvist}}},\ }\href
  {\doibase 10.1103/PhysRevLett.103.211101} {\bibfield  {journal} {\bibinfo
  {journal} {Phys.~Rev.~Lett.}\ }\textbf {\bibinfo {volume} {103}},\ \bibinfo
  {eid} {211101} (\bibinfo {year} {2009})},\ \Eprint
  {http://arxiv.org/abs/0908.1274} {arXiv:0908.1274 [astro-ph.SR]} \BibitemShut
  {NoStop}%
\bibitem [{\citenamefont {{Chen}}\ \emph {et~al.}(2016)\citenamefont {{Chen}},
  \citenamefont {{Matteini}}, \citenamefont {{Schekochihin}}, \citenamefont
  {{Stevens}}, \citenamefont {{Salem}}, \citenamefont {{Maruca}}, \citenamefont
  {{Kunz}},\ and\ \citenamefont {{Bale}}}]{Chen:2016}%
  \BibitemOpen
  \bibfield  {author} {\bibinfo {author} {\bibfnamefont {C.~H.~K.}\
  \bibnamefont {{Chen}}}, \bibinfo {author} {\bibfnamefont {L.}~\bibnamefont
  {{Matteini}}}, \bibinfo {author} {\bibfnamefont {A.~A.}\ \bibnamefont
  {{Schekochihin}}}, \bibinfo {author} {\bibfnamefont {M.~L.}\ \bibnamefont
  {{Stevens}}}, \bibinfo {author} {\bibfnamefont {C.~S.}\ \bibnamefont
  {{Salem}}}, \bibinfo {author} {\bibfnamefont {B.~A.}\ \bibnamefont
  {{Maruca}}}, \bibinfo {author} {\bibfnamefont {M.~W.}\ \bibnamefont
  {{Kunz}}}, \ and\ \bibinfo {author} {\bibfnamefont {S.~D.}\ \bibnamefont
  {{Bale}}},\ }\href {\doibase 10.3847/2041-8205/825/2/L26} {\bibfield
  {journal} {\bibinfo  {journal} {Astrophys.~J.~Lett.}\ }\textbf {\bibinfo
  {volume} {825}},\ \bibinfo {eid} {L26} (\bibinfo {year} {2016})},\ \Eprint
  {http://arxiv.org/abs/1606.02624} {arXiv:1606.02624 [physics.space-ph]}
  \BibitemShut {NoStop}%
\bibitem [{\citenamefont {Nyquist}(1932)}]{Nyquist:1932}%
  \BibitemOpen
  \bibfield  {author} {\bibinfo {author} {\bibfnamefont {H.}~\bibnamefont
  {Nyquist}},\ }\href@noop {} {\bibfield  {journal} {\bibinfo  {journal} {Bell
  system technical journal}\ }\textbf {\bibinfo {volume} {11}},\ \bibinfo
  {pages} {126} (\bibinfo {year} {1932})}\BibitemShut {NoStop}%
\bibitem [{\citenamefont {Klein}\ \emph {et~al.}(2017)\citenamefont {Klein},
  \citenamefont {Kasper}, \citenamefont {Korreck},\ and\ \citenamefont
  {Stevens}}]{Klein:2017c}%
  \BibitemOpen
  \bibfield  {author} {\bibinfo {author} {\bibfnamefont {K.~G.}\ \bibnamefont
  {Klein}}, \bibinfo {author} {\bibfnamefont {J.~C.}\ \bibnamefont {Kasper}},
  \bibinfo {author} {\bibfnamefont {K.~E.}\ \bibnamefont {Korreck}}, \ and\
  \bibinfo {author} {\bibfnamefont {M.~L.}\ \bibnamefont {Stevens}},\ }\href
  {\doibase 10.1002/2017JA024486} {\bibfield  {journal} {\bibinfo  {journal}
  {J.~Geophys.~Res.}\ ,\ \bibinfo {pages} {9815}} (\bibinfo {year} {2017})},\
  \bibinfo {note} {2017JA024486}\BibitemShut {NoStop}%
\bibitem [{\citenamefont {{Klein}}\ and\ \citenamefont
  {{Howes}}(2015)}]{Klein:2015a}%
  \BibitemOpen
  \bibfield  {author} {\bibinfo {author} {\bibfnamefont {K.~G.}\ \bibnamefont
  {{Klein}}}\ and\ \bibinfo {author} {\bibfnamefont {G.~G.}\ \bibnamefont
  {{Howes}}},\ }\href {\doibase 10.1063/1.4914933} {\bibfield  {journal}
  {\bibinfo  {journal} {Phys.~Plasmas}\ }\textbf {\bibinfo {volume} {22}},\
  \bibinfo {eid} {032903} (\bibinfo {year} {2015})},\ \Eprint
  {http://arxiv.org/abs/1503.00695} {arXiv:1503.00695 [physics.space-ph]}
  \BibitemShut {NoStop}%
\bibitem [{\citenamefont {{Gary}}\ \emph {et~al.}(2016)\citenamefont {{Gary}},
  \citenamefont {{Jian}}, \citenamefont {{Broiles}}, \citenamefont {{Stevens}},
  \citenamefont {{Podesta}},\ and\ \citenamefont {{Kasper}}}]{Gary:2016}%
  \BibitemOpen
  \bibfield  {author} {\bibinfo {author} {\bibfnamefont {S.~P.}\ \bibnamefont
  {{Gary}}}, \bibinfo {author} {\bibfnamefont {L.~K.}\ \bibnamefont {{Jian}}},
  \bibinfo {author} {\bibfnamefont {T.~W.}\ \bibnamefont {{Broiles}}}, \bibinfo
  {author} {\bibfnamefont {M.~L.}\ \bibnamefont {{Stevens}}}, \bibinfo {author}
  {\bibfnamefont {J.~J.}\ \bibnamefont {{Podesta}}}, \ and\ \bibinfo {author}
  {\bibfnamefont {J.~C.}\ \bibnamefont {{Kasper}}},\ }\href {\doibase
  10.1002/2015JA021935} {\bibfield  {journal} {\bibinfo  {journal}
  {J.~Geophys.~Res.}\ }\textbf {\bibinfo {volume} {121}},\ \bibinfo {pages}
  {30} (\bibinfo {year} {2016})}\BibitemShut {NoStop}%
\bibitem [{\citenamefont {{Ogilvie}}\ \emph {et~al.}(1995)\citenamefont
  {{Ogilvie}}, \citenamefont {{Chornay}}, \citenamefont {{Fritzenreiter}},
  \citenamefont {{Hunsaker}}, \citenamefont {{Keller}}, \citenamefont
  {{Lobell}}, \citenamefont {{Miller}}, \citenamefont {{Scudder}},
  \citenamefont {{Sittler}}, \citenamefont {{Torbert}}, \citenamefont
  {{Bodet}}, \citenamefont {{Needell}}, \citenamefont {{Lazarus}},
  \citenamefont {{Steinberg}}, \citenamefont {{Tappan}}, \citenamefont
  {{Mavretic}},\ and\ \citenamefont {{Gergin}}}]{Ogilvie:1995}%
  \BibitemOpen
  \bibfield  {author} {\bibinfo {author} {\bibfnamefont {K.~W.}\ \bibnamefont
  {{Ogilvie}}}, \bibinfo {author} {\bibfnamefont {D.~J.}\ \bibnamefont
  {{Chornay}}}, \bibinfo {author} {\bibfnamefont {R.~J.}\ \bibnamefont
  {{Fritzenreiter}}}, \bibinfo {author} {\bibfnamefont {F.}~\bibnamefont
  {{Hunsaker}}}, \bibinfo {author} {\bibfnamefont {J.}~\bibnamefont
  {{Keller}}}, \bibinfo {author} {\bibfnamefont {J.}~\bibnamefont {{Lobell}}},
  \bibinfo {author} {\bibfnamefont {G.}~\bibnamefont {{Miller}}}, \bibinfo
  {author} {\bibfnamefont {J.~D.}\ \bibnamefont {{Scudder}}}, \bibinfo {author}
  {\bibfnamefont {E.~C.}\ \bibnamefont {{Sittler}}, \bibfnamefont {Jr.}},
  \bibinfo {author} {\bibfnamefont {R.~B.}\ \bibnamefont {{Torbert}}}, \bibinfo
  {author} {\bibfnamefont {D.}~\bibnamefont {{Bodet}}}, \bibinfo {author}
  {\bibfnamefont {G.}~\bibnamefont {{Needell}}}, \bibinfo {author}
  {\bibfnamefont {A.~J.}\ \bibnamefont {{Lazarus}}}, \bibinfo {author}
  {\bibfnamefont {J.~T.}\ \bibnamefont {{Steinberg}}}, \bibinfo {author}
  {\bibfnamefont {J.~H.}\ \bibnamefont {{Tappan}}}, \bibinfo {author}
  {\bibfnamefont {A.}~\bibnamefont {{Mavretic}}}, \ and\ \bibinfo {author}
  {\bibfnamefont {E.}~\bibnamefont {{Gergin}}},\ }\href {\doibase
  10.1007/BF00751326} {\bibfield  {journal} {\bibinfo  {journal} {Space
  Sci.~Rev.}\ }\textbf {\bibinfo {volume} {71}},\ \bibinfo {pages} {55}
  (\bibinfo {year} {1995})}\BibitemShut {NoStop}%
\bibitem [{\citenamefont {Lepping}\ \emph {et~al.}(1995)\citenamefont
  {Lepping}, \citenamefont {Acũna}, \citenamefont {Burlaga}, \citenamefont
  {Farrell}, \citenamefont {Slavin}, \citenamefont {Schatten}, \citenamefont
  {Mariani}, \citenamefont {Ness}, \citenamefont {Neubauer}, \citenamefont
  {Whang}, \citenamefont {Byrnes}, \citenamefont {Kennon}, \citenamefont
  {Panetta}, \citenamefont {Scheifele},\ and\ \citenamefont
  {Worley}}]{Lepping:1995}%
  \BibitemOpen
  \bibfield  {author} {\bibinfo {author} {\bibfnamefont {R.~P.}\ \bibnamefont
  {Lepping}}, \bibinfo {author} {\bibfnamefont {M.~H.}\ \bibnamefont {Acũna}},
  \bibinfo {author} {\bibfnamefont {L.~F.}\ \bibnamefont {Burlaga}}, \bibinfo
  {author} {\bibfnamefont {W.~M.}\ \bibnamefont {Farrell}}, \bibinfo {author}
  {\bibfnamefont {J.~A.}\ \bibnamefont {Slavin}}, \bibinfo {author}
  {\bibfnamefont {K.~H.}\ \bibnamefont {Schatten}}, \bibinfo {author}
  {\bibfnamefont {F.}~\bibnamefont {Mariani}}, \bibinfo {author} {\bibfnamefont
  {N.~F.}\ \bibnamefont {Ness}}, \bibinfo {author} {\bibfnamefont {F.~M.}\
  \bibnamefont {Neubauer}}, \bibinfo {author} {\bibfnamefont {Y.~C.}\
  \bibnamefont {Whang}}, \bibinfo {author} {\bibfnamefont {J.~B.}\ \bibnamefont
  {Byrnes}}, \bibinfo {author} {\bibfnamefont {R.~S.}\ \bibnamefont {Kennon}},
  \bibinfo {author} {\bibfnamefont {P.~V.}\ \bibnamefont {Panetta}}, \bibinfo
  {author} {\bibfnamefont {J.}~\bibnamefont {Scheifele}}, \ and\ \bibinfo
  {author} {\bibfnamefont {E.~M.}\ \bibnamefont {Worley}},\ }\href
  {http://dx.doi.org/10.1007/BF00751330} {\bibfield  {journal} {\bibinfo
  {journal} {Space Sci.~Rev.}\ }\textbf {\bibinfo {volume} {71}},\ \bibinfo
  {pages} {207} (\bibinfo {year} {1995})},\ \bibinfo {note}
  {10.1007/BF00751330}\BibitemShut {NoStop}%
\bibitem [{\citenamefont {{Koval}}\ and\ \citenamefont
  {{Szabo}}(2013)}]{Koval:2013}%
  \BibitemOpen
  \bibfield  {author} {\bibinfo {author} {\bibfnamefont {A.}~\bibnamefont
  {{Koval}}}\ and\ \bibinfo {author} {\bibfnamefont {A.}~\bibnamefont
  {{Szabo}}},\ }\href {\doibase 10.1063/1.4811025} {\bibfield  {journal}
  {\bibinfo  {journal} {Solar Wind 13}\ }\textbf {\bibinfo {volume} {1539}},\
  \bibinfo {pages} {211} (\bibinfo {year} {2013})}\BibitemShut {NoStop}%
\bibitem [{\citenamefont {{Kunz}}\ \emph {et~al.}(2017)\citenamefont {{Kunz}},
  \citenamefont {{Abel}}, \citenamefont {{Klein}},\ and\ \citenamefont
  {{Schekochihin}}}]{Kunz:2018}%
  \BibitemOpen
  \bibfield  {author} {\bibinfo {author} {\bibfnamefont {M.~W.}\ \bibnamefont
  {{Kunz}}}, \bibinfo {author} {\bibfnamefont {I.~G.}\ \bibnamefont {{Abel}}},
  \bibinfo {author} {\bibfnamefont {K.~G.}\ \bibnamefont {{Klein}}}, \ and\
  \bibinfo {author} {\bibfnamefont {A.~A.}\ \bibnamefont {{Schekochihin}}},\
  }\href@noop {} {\bibfield  {journal} {\bibinfo  {journal} {ArXiv e-prints}\ }
  (\bibinfo {year} {2017})},\ \Eprint {http://arxiv.org/abs/1712.02269}
  {arXiv:1712.02269 [astro-ph.HE]} \BibitemShut {NoStop}%
\bibitem [{\citenamefont {{Wilson}}\ \emph {et~al.}(2018)\citenamefont
  {{Wilson}}, \citenamefont {{Stevens}}, \citenamefont {{Kasper}},
  \citenamefont {{Klein}}, \citenamefont {{Maruca}}, \citenamefont {{Bale}},
  \citenamefont {{Bowen}}, \citenamefont {{Pulupa}},\ and\ \citenamefont
  {{Salem}}}]{Wilson:2018}%
  \BibitemOpen
  \bibfield  {author} {\bibinfo {author} {\bibfnamefont {L.~B.}\ \bibnamefont
  {{Wilson}}, \bibfnamefont {III}}, \bibinfo {author} {\bibfnamefont {M.~L.}\
  \bibnamefont {{Stevens}}}, \bibinfo {author} {\bibfnamefont {J.~C.}\
  \bibnamefont {{Kasper}}}, \bibinfo {author} {\bibfnamefont {K.~G.}\
  \bibnamefont {{Klein}}}, \bibinfo {author} {\bibfnamefont {B.~A.}\
  \bibnamefont {{Maruca}}}, \bibinfo {author} {\bibfnamefont {S.~D.}\
  \bibnamefont {{Bale}}}, \bibinfo {author} {\bibfnamefont {T.~A.}\
  \bibnamefont {{Bowen}}}, \bibinfo {author} {\bibfnamefont {M.~P.}\
  \bibnamefont {{Pulupa}}}, \ and\ \bibinfo {author} {\bibfnamefont {C.~S.}\
  \bibnamefont {{Salem}}},\ }\href@noop {} {\bibfield  {journal} {\bibinfo
  {journal} {ArXiv e-prints}\ } (\bibinfo {year} {2018})},\ \Eprint
  {http://arxiv.org/abs/1802.08585} {arXiv:1802.08585 [physics.plasm-ph]}
  \BibitemShut {NoStop}%
\bibitem [{\citenamefont {{Hellinger}}(2007)}]{Hellinger:2007}%
  \BibitemOpen
  \bibfield  {author} {\bibinfo {author} {\bibfnamefont {P.}~\bibnamefont
  {{Hellinger}}},\ }\href {\doibase 10.1063/1.2768318} {\bibfield  {journal}
  {\bibinfo  {journal} {Phys.~Plasmas}\ }\textbf {\bibinfo {volume} {14}},\
  \bibinfo {eid} {082105} (\bibinfo {year} {2007})}\BibitemShut {NoStop}%
\bibitem [{\citenamefont {{Kunz}}\ \emph {et~al.}(2015)\citenamefont {{Kunz}},
  \citenamefont {{Schekochihin}}, \citenamefont {{Chen}}, \citenamefont
  {{Abel}},\ and\ \citenamefont {{Cowley}}}]{Kunz:2015}%
  \BibitemOpen
  \bibfield  {author} {\bibinfo {author} {\bibfnamefont {M.~W.}\ \bibnamefont
  {{Kunz}}}, \bibinfo {author} {\bibfnamefont {A.~A.}\ \bibnamefont
  {{Schekochihin}}}, \bibinfo {author} {\bibfnamefont {C.~H.~K.}\ \bibnamefont
  {{Chen}}}, \bibinfo {author} {\bibfnamefont {I.~G.}\ \bibnamefont {{Abel}}},
  \ and\ \bibinfo {author} {\bibfnamefont {S.~C.}\ \bibnamefont {{Cowley}}},\
  }\href {\doibase 10.1017/S0022377815000811} {\bibfield  {journal} {\bibinfo
  {journal} {J.~Plasma Phys.}\ }\textbf {\bibinfo {volume} {81}},\ \bibinfo
  {eid} {325810501} (\bibinfo {year} {2015})},\ \Eprint
  {http://arxiv.org/abs/1501.06771} {arXiv:1501.06771 [astro-ph.HE]}
  \BibitemShut {NoStop}%
\bibitem [{\citenamefont {{Goldreich}}\ and\ \citenamefont
  {{Sridhar}}(1995)}]{Goldreich:1995}%
  \BibitemOpen
  \bibfield  {author} {\bibinfo {author} {\bibfnamefont {P.}~\bibnamefont
  {{Goldreich}}}\ and\ \bibinfo {author} {\bibfnamefont {S.}~\bibnamefont
  {{Sridhar}}},\ }\href {\doibase 10.1086/175121} {\bibfield  {journal}
  {\bibinfo  {journal} {Astrophys.~J.}\ }\textbf {\bibinfo {volume} {438}},\
  \bibinfo {pages} {763} (\bibinfo {year} {1995})}\BibitemShut {NoStop}%
\bibitem [{\citenamefont {{Mallet}}\ \emph {et~al.}(2015)\citenamefont
  {{Mallet}}, \citenamefont {{Schekochihin}},\ and\ \citenamefont
  {{Chandran}}}]{Mallet:2015}%
  \BibitemOpen
  \bibfield  {author} {\bibinfo {author} {\bibfnamefont {A.}~\bibnamefont
  {{Mallet}}}, \bibinfo {author} {\bibfnamefont {A.~A.}\ \bibnamefont
  {{Schekochihin}}}, \ and\ \bibinfo {author} {\bibfnamefont {B.~D.~G.}\
  \bibnamefont {{Chandran}}},\ }\href {\doibase 10.1093/mnrasl/slv021}
  {\bibfield  {journal} {\bibinfo  {journal} {Mon.~Not.~Roy.~Astron.~Soc.}\
  }\textbf {\bibinfo {volume} {449}},\ \bibinfo {pages} {L77} (\bibinfo {year}
  {2015})},\ \Eprint {http://arxiv.org/abs/1406.5658} {arXiv:1406.5658
  [astro-ph.SR]} \BibitemShut {NoStop}%
\bibitem [{\citenamefont {{Kasper}}\ \emph {et~al.}(2006)\citenamefont
  {{Kasper}}, \citenamefont {{Lazarus}}, \citenamefont {{Steinberg}},
  \citenamefont {{Ogilvie}},\ and\ \citenamefont {{Szabo}}}]{Kasper:2006}%
  \BibitemOpen
  \bibfield  {author} {\bibinfo {author} {\bibfnamefont {J.~C.}\ \bibnamefont
  {{Kasper}}}, \bibinfo {author} {\bibfnamefont {A.~J.}\ \bibnamefont
  {{Lazarus}}}, \bibinfo {author} {\bibfnamefont {J.~T.}\ \bibnamefont
  {{Steinberg}}}, \bibinfo {author} {\bibfnamefont {K.~W.}\ \bibnamefont
  {{Ogilvie}}}, \ and\ \bibinfo {author} {\bibfnamefont {A.}~\bibnamefont
  {{Szabo}}},\ }\href {\doibase 10.1029/2005JA011442} {\bibfield  {journal}
  {\bibinfo  {journal} {Journal of Geophysical Research (Space Physics)}\
  }\textbf {\bibinfo {volume} {111}},\ \bibinfo {eid} {A03105} (\bibinfo {year}
  {2006})}\BibitemShut {NoStop}%
\bibitem [{\citenamefont {{Dum}}\ \emph {et~al.}(1980)\citenamefont {{Dum}},
  \citenamefont {{Marsch}},\ and\ \citenamefont {{Pilipp}}}]{Dum:1980}%
  \BibitemOpen
  \bibfield  {author} {\bibinfo {author} {\bibfnamefont {C.~T.}\ \bibnamefont
  {{Dum}}}, \bibinfo {author} {\bibfnamefont {E.}~\bibnamefont {{Marsch}}}, \
  and\ \bibinfo {author} {\bibfnamefont {W.}~\bibnamefont {{Pilipp}}},\ }\href
  {\doibase 10.1017/S0022377800022170} {\bibfield  {journal} {\bibinfo
  {journal} {J.~Plasma Phys.}\ }\textbf {\bibinfo {volume} {23}},\ \bibinfo
  {pages} {91} (\bibinfo {year} {1980})}\BibitemShut {NoStop}%
\bibitem [{\citenamefont {{Isenberg}}(2012)}]{Isenberg:2012}%
  \BibitemOpen
  \bibfield  {author} {\bibinfo {author} {\bibfnamefont {P.~A.}\ \bibnamefont
  {{Isenberg}}},\ }\href {\doibase 10.1063/1.3697721} {\bibfield  {journal}
  {\bibinfo  {journal} {Phys.~Plasmas}\ }\textbf {\bibinfo {volume} {19}},\
  \bibinfo {pages} {032116} (\bibinfo {year} {2012})},\ \Eprint
  {http://arxiv.org/abs/1203.1938} {arXiv:1203.1938 [physics.plasm-ph]}
  \BibitemShut {NoStop}%
\bibitem [{\citenamefont {{Verscharen}}\ \emph {et~al.}(view)\citenamefont
  {{Verscharen}}, \citenamefont {{Klein}}, \citenamefont {{Chandran}},
  \citenamefont {{Stevens}}, \citenamefont {{Salem}},\ and\ \citenamefont
  {{Bale}}}]{Verscharen:2018}%
  \BibitemOpen
  \bibfield  {author} {\bibinfo {author} {\bibfnamefont {D.}~\bibnamefont
  {{Verscharen}}}, \bibinfo {author} {\bibfnamefont {K.~G.}\ \bibnamefont
  {{Klein}}}, \bibinfo {author} {\bibfnamefont {B.~D.~G.}\ \bibnamefont
  {{Chandran}}}, \bibinfo {author} {\bibfnamefont {M.~L.}\ \bibnamefont
  {{Stevens}}}, \bibinfo {author} {\bibfnamefont {C.~S.}\ \bibnamefont
  {{Salem}}}, \ and\ \bibinfo {author} {\bibfnamefont {S.~D.}\ \bibnamefont
  {{Bale}}},\ }\href@noop {} {\bibfield  {journal} {\bibinfo  {journal}
  {J.~Plasma Phys.}\ } (\bibinfo {year} {under review})}\BibitemShut {NoStop}%
\bibitem [{\citenamefont {{Fox}}\ \emph {et~al.}(2016)\citenamefont {{Fox}},
  \citenamefont {{Velli}}, \citenamefont {{Bale}}, \citenamefont {{Decker}},
  \citenamefont {{Driesman}}, \citenamefont {{Howard}}, \citenamefont
  {{Kasper}}, \citenamefont {{Kinnison}}, \citenamefont {{Kusterer}},
  \citenamefont {{Lario}}, \citenamefont {{Lockwood}}, \citenamefont
  {{McComas}}, \citenamefont {{Raouafi}},\ and\ \citenamefont
  {{Szabo}}}]{Fox:2016}%
  \BibitemOpen
  \bibfield  {author} {\bibinfo {author} {\bibfnamefont {N.~J.}\ \bibnamefont
  {{Fox}}}, \bibinfo {author} {\bibfnamefont {M.~C.}\ \bibnamefont {{Velli}}},
  \bibinfo {author} {\bibfnamefont {S.~D.}\ \bibnamefont {{Bale}}}, \bibinfo
  {author} {\bibfnamefont {R.}~\bibnamefont {{Decker}}}, \bibinfo {author}
  {\bibfnamefont {A.}~\bibnamefont {{Driesman}}}, \bibinfo {author}
  {\bibfnamefont {R.~A.}\ \bibnamefont {{Howard}}}, \bibinfo {author}
  {\bibfnamefont {J.~C.}\ \bibnamefont {{Kasper}}}, \bibinfo {author}
  {\bibfnamefont {J.}~\bibnamefont {{Kinnison}}}, \bibinfo {author}
  {\bibfnamefont {M.}~\bibnamefont {{Kusterer}}}, \bibinfo {author}
  {\bibfnamefont {D.}~\bibnamefont {{Lario}}}, \bibinfo {author} {\bibfnamefont
  {M.~K.}\ \bibnamefont {{Lockwood}}}, \bibinfo {author} {\bibfnamefont
  {D.~J.}\ \bibnamefont {{McComas}}}, \bibinfo {author} {\bibfnamefont {N.~E.}\
  \bibnamefont {{Raouafi}}}, \ and\ \bibinfo {author} {\bibfnamefont
  {A.}~\bibnamefont {{Szabo}}},\ }\href {\doibase 10.1007/s11214-015-0211-6}
  {\bibfield  {journal} {\bibinfo  {journal} {Space Sci.~Rev.}\ }\textbf
  {\bibinfo {volume} {204}},\ \bibinfo {pages} {7} (\bibinfo {year}
  {2016})}\BibitemShut {NoStop}%
\bibitem [{\citenamefont {{M{\"u}ller}}\ \emph {et~al.}(2013)\citenamefont
  {{M{\"u}ller}}, \citenamefont {{Marsden}}, \citenamefont {{St.~Cyr}},\ and\
  \citenamefont {{Gilbert}}}]{Muller:2013}%
  \BibitemOpen
  \bibfield  {author} {\bibinfo {author} {\bibfnamefont {D.}~\bibnamefont
  {{M{\"u}ller}}}, \bibinfo {author} {\bibfnamefont {R.~G.}\ \bibnamefont
  {{Marsden}}}, \bibinfo {author} {\bibfnamefont {O.~C.}\ \bibnamefont
  {{St.~Cyr}}}, \ and\ \bibinfo {author} {\bibfnamefont {H.~R.}\ \bibnamefont
  {{Gilbert}}},\ }\href {\doibase 10.1007/s11207-012-0085-7} {\bibfield
  {journal} {\bibinfo  {journal} {Sol.~Phys.}\ }\textbf {\bibinfo {volume}
  {285}},\ \bibinfo {pages} {25} (\bibinfo {year} {2013})},\ \Eprint
  {http://arxiv.org/abs/1207.4579} {arXiv:1207.4579 [astro-ph.SR]} \BibitemShut
  {NoStop}%
\end{thebibliography}

%

\end{document}